\documentclass[aps,prl,reprint,superscriptaddress]{revtex4-1}

\usepackage{amsmath}
\usepackage{amssymb}
\usepackage{graphicx}
\usepackage{relsize}

\pdfoutput=1

\newcommand{\eq}{\begin{equation}}
\newcommand{\eeq}{\end{equation}}
\newcommand{\eqarray}{\begin{eqnarray}}
\newcommand{\eeqarray}{\end{eqnarray}}
\newcommand{\szsubspace}{\mathcal{S}_z=0}
\newcommand{\ket}[1]{\left|#1\right\rangle}
\newcommand{\bra}[1]{\left\langle #1\right|}
\newcommand{\adag}{a^\dagger}

\graphicspath{{./images/}}

\newcommand{\FigureOne}{
\begin{figure}
\includegraphics[width=\columnwidth]{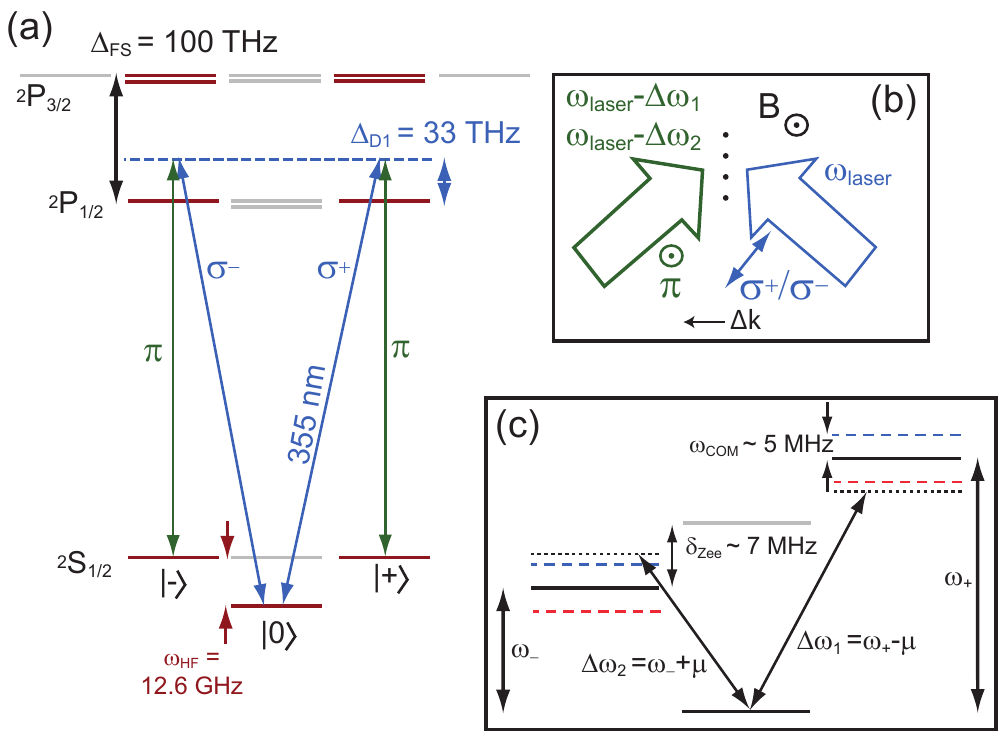}
\caption{(a): Level diagram for $^{171}$Yb$^+$, highlighting relevant states. (b): Sketch of experimental geometry, showing the directions of the laser wavevectors and the real magnetic field relative to the ion chain. Both beams are linearly polarized, one along the $\vec{B}$ field (providing $\pi$ light) and one orthogonal to the $\vec{B}$ field (providing an equal superposition of $\sigma^+$ and $\sigma^-$ light). Multiple beatnotes are applied by imprinting multiple frequencies onto one beam (in this case, the $\pi$-polarized beam). (c): Detailed level diagram of the $^2$S$_{1/2}$ ground state, showing Raman beatnotes in relation to Zeeman splittings and motional sidebands for the center-of-mass mode. Level splittings are not to scale. }
\label{fig:spin1levels}
\end{figure}
}

\newcommand{\FigureTwo}{
\begin{figure}
\includegraphics[width=\columnwidth]{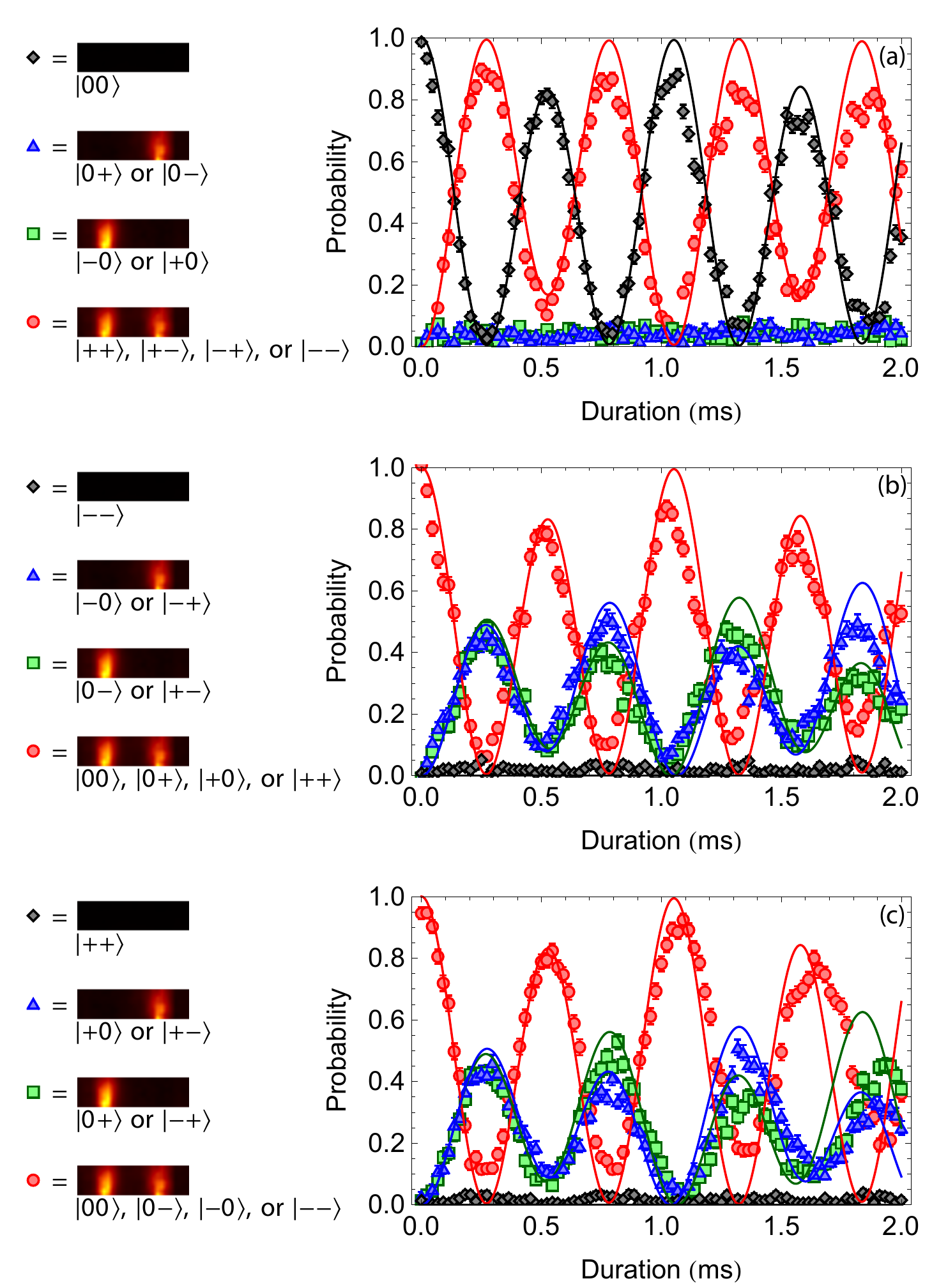}
\caption{Dynamics of 2 spin-1 particles evolving under the XY Hamiltonian in Eq. \ref{eq:spin1XY}. (a) We measure the populations for each ion to be in the state $\ket{0}$. The probability of both ions to be in $\ket{0}$ (black diamonds), only the left ion in $\ket{0}$ (blue triangles), only the right ion in $\ket{0}$, and neither ion in $\ket{0}$ (red circles) are plotted together. Similar plots for the $\ket{-}$ state (b) and $\ket{+}$ state (c) are also shown. The dynamics resemble Rabi flopping between the state $\ket{00}$ and the symmetric superposition $(\ket{+-} + \ket{-+})/\sqrt{2}$. Solid lines represent theoretical predictions, with the only free parameters being the magnitude of the $S_z$ and $(S_z)^2$ gradients discussed in the text. In panel (c), the interaction $J_{1,2}$ drifts at long times compared to that estimated from Eq. \ref{eq:Jij}, consistent with an observed drift in radial trap frequencies during the data collection. Error bars (1 s.d.) show the statistical uncertainty based on 500 repetitions of the experiment.}
\label{fig:dynamics2ions}
\end{figure}
}

\newcommand{\FigureThree}{
\begin{figure}
\includegraphics[scale=0.9]{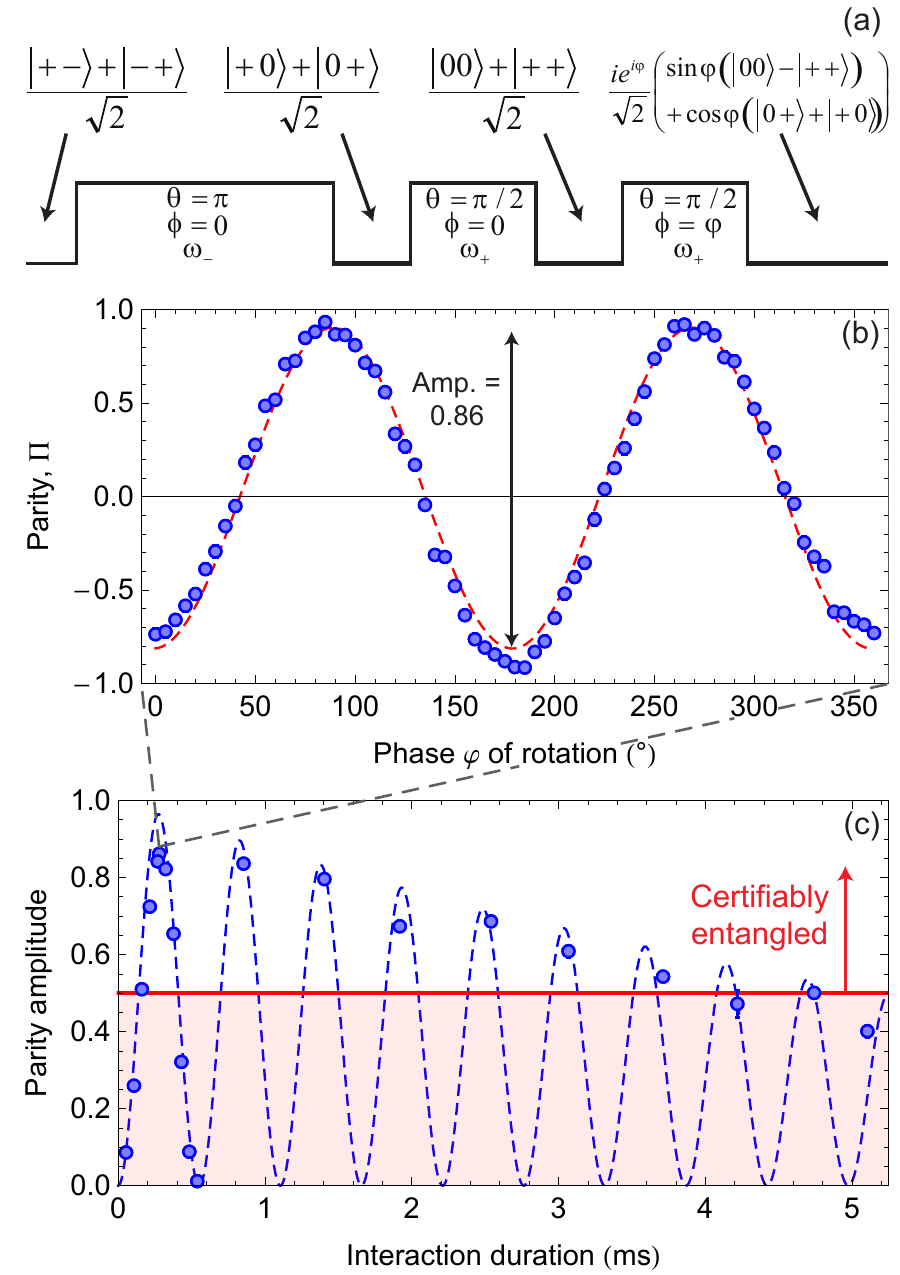}
\caption{(a): Illustration of the rotations performed before measuring the parity, where rotations by $\theta$ and $\varphi$ are defined in Eq. \ref{eqn:Rotations}. Also shown are the ideal initial state and the states produced after each step. (b): The parity of the final state oscillates as a function of the final pulse phase $\varphi$. Fitting the function $C - A\cos2\varphi + B \sin2\varphi$ (red dashed line) to the data results in an amplitude $A = 0.86 >0.5$, demonstrating entanglement. (c): Amplitude of the parity oscillation after various durations of the spin-spin interaction, showing the peak amplitude for each time $\sim(2n+1)/(2\sqrt{2}J_{1,2})$. The data in (b) correspond to the highest-contrast point in (c). The dashed line is a guide to the eye, suggesting the expected behavior of sinusoidal oscillations between a product state and an entangled state, along with decay due to decoherence at longer times. The chosen durations are not evenly spaced past 0.6 ms: for each point, the duration was chosen such that the population in $\ket{00}$ was at a local minimum, and drifts in the radial trap frequencies led to small changes in $J_{1,2}$. }
\label{fig:ParityFig}
\end{figure}
}

\newcommand{\FigureFour}{
\begin{figure}
\includegraphics[width=\columnwidth]{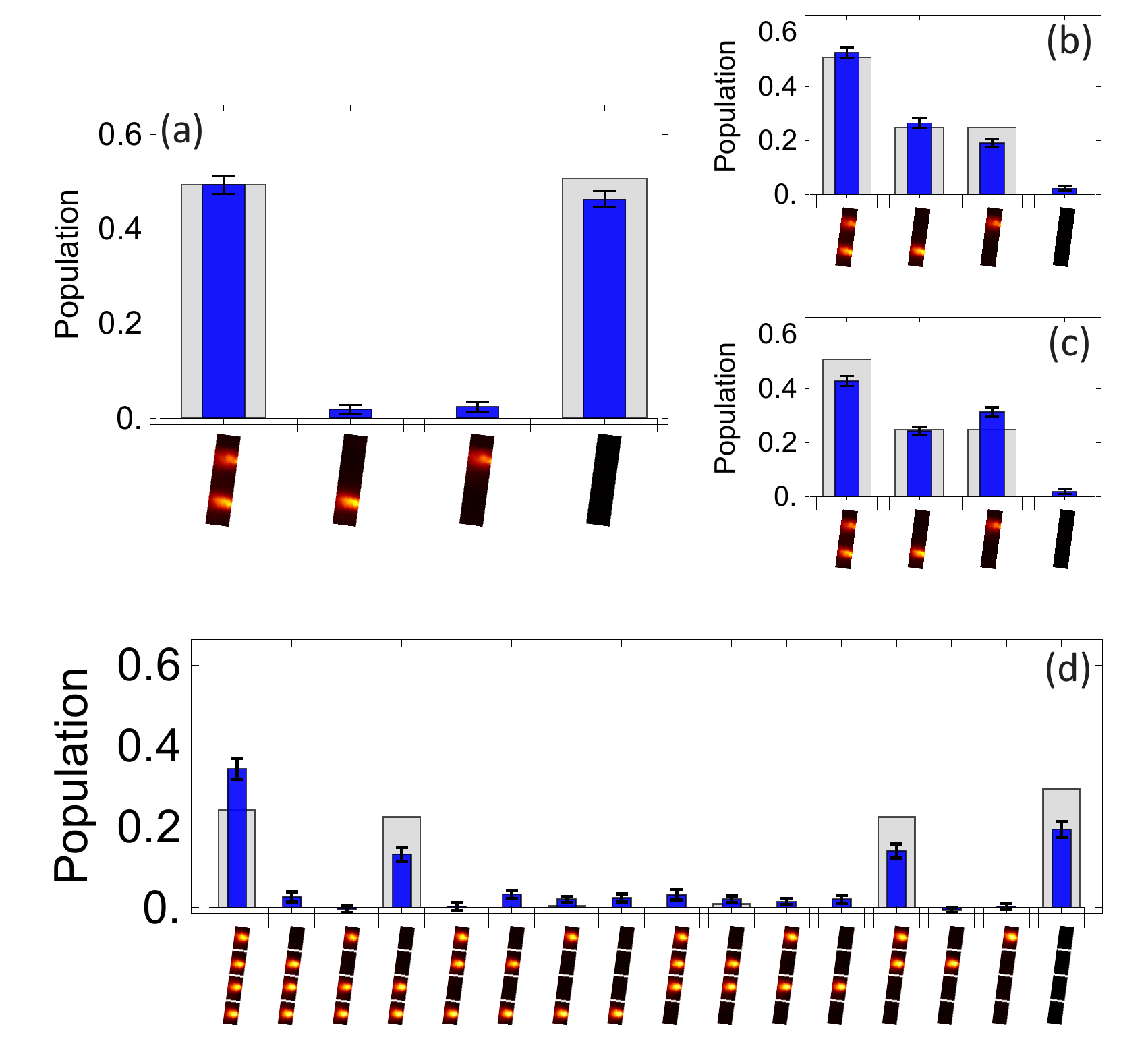}
\caption{Measurements of the prepared 2-spin (a-c) and 4-spin (d) states after ramping an $(S_z)^2$ field (narrow blue bars) compared to the values expected for the calculated ground state (gray bars). As in Fig. \ref{fig:dynamics2ions}, panels (a), (b), and (c) show the measured populations when the `dark' state is set to be $\ket{0}$, $\ket{-}$, or $\ket{+}$, respectively. The dark state is set to $\ket{0}$ in part (d).}
\label{fig:EvenIonGroundState}
\end{figure}
}

\newcommand{\FigureFive}{
\begin{figure}
\includegraphics[width=\columnwidth]{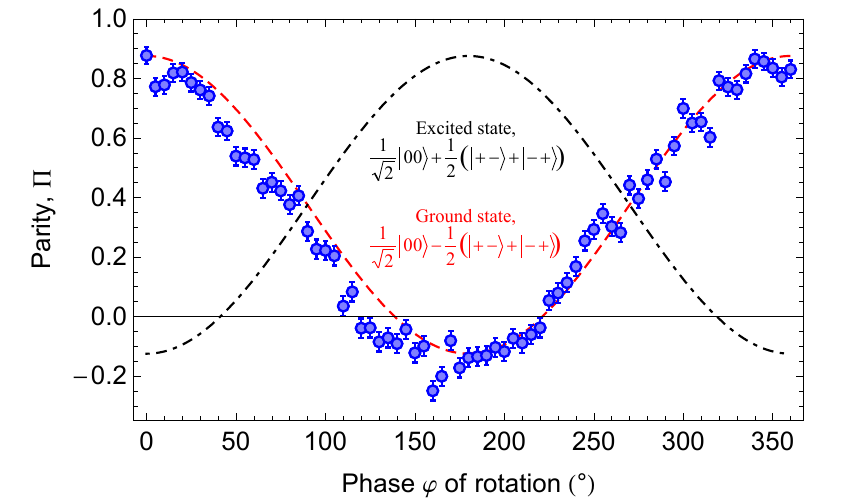}
\caption{Following an adiabatic ramp, the parity of the final state is measured as a function of the final rotation phase $\varphi$ (see text for rotation protocol). Dashed and dot-dashed lines represent the theoretically expected values for the ground state, $\ket{00}/\sqrt{2} - \left( \ket{-+} + \ket{+-} \right)/2$, and highest excited state, $\ket{00}/\sqrt{2} + \left( \ket{-+} + \ket{+-} \right)/2$, respectively. The phase of the oscillation reveals that the relative phases in the prepared state are consistent with the expected ground state. }
\label{fig:GroundStatePhase}
\end{figure}
}

\newcommand{\FigureSix}{
\begin{figure}
\includegraphics[width=\columnwidth]{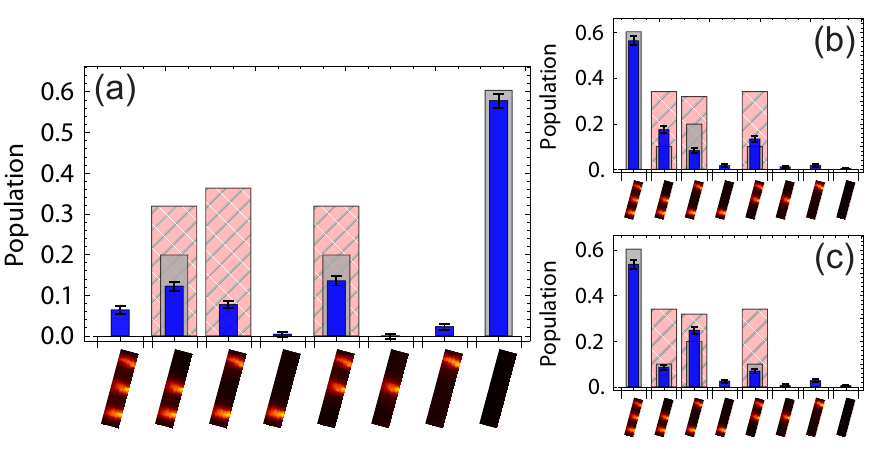}
\caption{Measurements when the $\ket{0}$ state is set dark (a), the $\ket{+}$ state is dark (b), and the $\ket{-}$ state is dark (c), of the prepared 3-spin state after adiabatically ramping a global $(S_z^i)^2$ field (narrow blue bars). The data agree closely with the calculated populations in the first excited state (gray bars), while showing little overlap with the expected populations in the ground state (wide, hatched red bars).}
\label{fig:3ionGroundState}
\end{figure}
}

\begin{document}

\title{Realization of a Quantum Integer-Spin Chain with Controllable Interactions}

\author{C. Senko$^1$, P. Richerme$^1$, J. Smith$^1$, A. Lee}
\affiliation{Joint Quantum Institute, University of Maryland Department of Physics and National Institute of Standards and Technology, College Park, MD 20742}

\author{I. Cohen$^2$, A. Retzker}
\affiliation{Racah Institute of Physics, The Hebrew University of Jerusalem, Jerusalem 91904, Givat Ram, Israel}

\author{C. Monroe}
\affiliation{Joint Quantum Institute, University of Maryland Department of Physics and National Institute of Standards and Technology, College Park, MD 20742}

\date{\today}

\begin{abstract}

The physics of interacting integer-spin chains has been a topic of intense theoretical interest, particularly in the context of symmetry-protected topological phases. However, there has not been a controllable model system to study this physics experimentally. We demonstrate how spin-dependent forces on trapped ions can be used to engineer an effective system of interacting \mbox{spin-1} particles. Our system evolves coherently under an applied spin-1 XY Hamiltonian with tunable, long-range couplings, and all three quantum levels at each site participate in the dynamics. We observe the time evolution of the system and verify its coherence by entangling a pair of effective three-level particles (`qutrits') with $86\%$ fidelity. By adiabatically ramping a global field, we produce ground states of the XY model, and we demonstrate an instance where the ground state cannot be created without breaking the same symmetries that protect the topological Haldane phase. This experimental platform enables future studies of symmetry-protected order in spin-1 systems and their use in quantum applications.
\end{abstract}

\maketitle

\section{Introduction}
A major area of current research is devoted to developing experimentally controllable systems that can be used for quantum computation, quantum communication, and quantum simulation of many-body physics. To date, most experiments have focused on the use of two-level systems (`qubits') for computation and communication \cite{Scarani2009,Ladd2010} and for the study of spin-1/2 (or spinless) many-body phenomena \cite{Bloch2012,Blatt2012}. However, there are a variety of motivations for performing experiments in higher-dimensional Hilbert spaces. Contrary to the intuition that enlarging the spin degree simplifies calculations by making them semiclassical \cite{Auerbach1994}, spin $> 1/2$ systems inherently have more complexity and cost exponentially more resources to classically simulate. For instance, it is computationally easy to find the ground state energy of a spin-1/2 chain with nearest-neighbor-only interactions in one-dimension; for systems with spin-7/2 or higher, the problem is known to belong to the QMA-complete complexity class, which is a quantum analogue of the classical NP-complete class \cite{Aharonov2007,HallgrenArxiv}. The difficulty of this problem for intermediate spin values, such as spin-1, is still an open question. From a more practical point of view, controllable three-level systems (`qutrits') are useful for quantum logic, since they can substantially simplify certain operations within quantum algorithms \cite{Lanyon2009} and can enhance the efficiency of quantum communication protocols \cite{Brukner2002}. 

When individual three-level systems are coupled together, they can be used to encode the physics of interacting \mbox{spin-1} particles. Such systems have attracted a great deal of theoretical interest following Haldane's conjecture that antiferromagnetic Heisenberg spin-1 chains, as opposed to spin-1/2 systems, have a finite energy gap that corresponds to exponentially decaying correlation functions \cite{Haldane1983,Haldane1983a}. This so-called Haldane phase possesses a doubly-degenerate entanglement spectrum \cite{Pollmann2010} and a non-local string order \cite{Kennedy1992,Kennedy1992a}, which is related to the order appearing in spin liquids \cite{Balents2010} and in the fractional quantum hall effect. These characteristics suggest that the Haldane phase is one of the simplest known examples of a symmetry-protected topological phase of matter \cite{Kennedy1992}. 

In addition to their interesting many-body properties, topological phases may be exploited in a more applied setting. The Haldane phase is useful for quantum operations (for instance, as a perfect quantum wire) \cite{Darmawan2010,Asoudeh2013} and can only be destroyed by crossing a phase transition. The finite energy gap in topological spin-1 systems makes them a potential candidate for long-lived, robust quantum memories \cite{Else2012NJP}, and schemes using symmetry-protected \mbox{spin-1} phases for measurement-based quantum computation have also been proposed \cite{Brennen2008,Else2012PRL}.

Several groups have developed controllable three-level quantum systems by using pairs of photons \cite{Lanyon2008} or superconducting circuits \cite{Bianchetti2010} to implement qutrits, or by using spinor BECs to study quantum magnetism \cite{Chang2005,Stamper-Kurn2013,Parker2013}. However, no platform has yet used multiple interacting qutrits for quantum information protocols or for simulating lattice spin models. In this paper, we use trapped atomic ions to simulate a chain of \mbox{spin-1} particles with tunable, long-range XY interactions \cite{Cohen2014}. Our system performs the same basic tasks that are commonly used in spin-1/2 quantum simulations, such as observing dynamical state evolution \cite{Richerme2014}, measuring coherence and certifying entanglement \cite{Kim2010}, and adiabatically preparing nontrivial ground states \cite{Richerme2013}. With two spin-1 particles, we observe coherent evolution under the XY interactions among states in a `decoherence free' subspace \cite{Lidar1998,Kielpinski2001}. For certain states generated by the XY Hamiltonian, we can verify entanglement between a pair of 3-level systems with fidelities of up to 86\%. Adding a time dependent global field allows us to adiabatically prepare the ground state of the XY model for even numbers of spins. For odd numbers of spins, producing the calculated ground state is not possible with a simple adiabatic ramp since it requires crossing a first-order phase transition, hinting at the existence of a symmetry-protected phase. The tools demonstrated here could enable future studies of symmetry-protected order and can be extended to SU(3) models and other systems of higher symmetry \cite{Grass2013}.

\FigureOne

\section{Experimental implementation}

The spin-1 chain is represented by a string of  $^{171}$Yb$^+$ atoms held in a linear Paul trap. Three hyperfine levels in the $^2S_{1/2}$ ground manifold of each atom are used to encode the spin-1 states: $\ket{+} \equiv \ket{F=1,m_F=1}$, $\ket{-} \equiv \ket{F=1,m_F=-1}$, and $\ket{0} \equiv \ket{F=0,m_F=0}$, with frequency splittings of $\omega_\pm$ between the $\ket{0}$ and $\ket{\pm}$ states, as shown in Fig. \ref{fig:spin1levels}. Here, $\ket{+}$, $\ket{-}$, and $\ket{0}$ are the eigenstates of $S_z$ with eigenvalues +1, -1, and 0 respectively; $F$ and $m_F$ are quantum numbers associated with the total angular momentum of the atom and its projection along the quantization axis, defined by a magnetic field of $\sim$5 G. 

We apply global laser beams to the ion chain with a wavevector difference along a principal axis of transverse motion, driving stimulated Raman transitions between the $\ket{0}$ and $\ket{-}$ states and between the $\ket{0}$ and $\ket{+}$ states with equal Rabi frequencies $\Omega_i$ on ion $i$ \cite{Kim2009}. To generate spin-1 XY interactions, we apply two beat frequencies at $\omega_- + \mu$ and $\omega_+ - \mu$ to these respective transitions, where $\mu$ is slightly detuned from the transverse motional frequencies, as shown in Fig. 1(c). Under the rotating wave approximations $\omega_\pm \gg \mu \gg \Omega_i$ and within the Lamb-Dicke regime ($\Delta k \left< \hat{x}_i \right> \ll 1$, with $\Delta k$ the wavevector difference of the Raman beams and $\hat{x}_i$ the position operator of the $i$th ion), the resulting interaction Hamiltonian (with $h=1$) is
\eq
H = \sum_{i,m=1}^N \frac{i\eta_{i,m}\Omega_i}{2\sqrt{2}} \left( -S^i_+ a_m e^{i(\mu-\omega_m) t} + S^i_- \adag_m e^{-i(\mu-\omega_m) t} \right).
\eeq
Here $a_m$ and $\adag_m$ are the phonon operators of the normal mode $m$ with frequency $\omega_m$, $\eta_{i,m} = b_{i,m}\sqrt{\hbar (\Delta k)^2/2 M \omega_m}$ is the Lamb-Dicke factor (where $b_{i,m}$ is the normal mode transformation matrix \cite{James1998} and $M$ is the mass of a single ion), and the spin raising and lowering operators $S^i_\pm$ satisfy the commutation relations $\left[S_+^i, S_-^j\right] = 2 S_z^i \delta_{ij}$. In the limit where the beatnotes are far detuned ($\eta_{i,m} \Omega_i \ll |\mu - \omega_m|$) and the phonons are only virtually excited, this results in an effective Hamiltonian with XY-type spin-spin interactions and spin-phonon couplings,
\begin{eqnarray}
\nonumber
H_{\mathrm{eff}} = &\sum_{i<j}& \frac{J_{i,j}}{4} \left(S_+^i S_-^j + S_-^i S_+^j \right) \\
\label{eqn:XYHam}
+ &\sum_{i,m}& V_{i,m} \left[ \left(2 a_m^\dagger a_m +1 \right) S_z^i - \left( S_z^i \right)^2 \right].
\label{eq:spin1XY}
\end{eqnarray}
 
The pure spin-spin interaction in the first term of Eq. \ref{eq:spin1XY} follows the same formula as for generating spin-1/2 Ising interactions \cite{Kim2009}:
\eq
J_{i,j} = \Omega_i \Omega_j \sum_m \frac{\eta_{i,m} \eta_{j,m}}{2(\mu-\omega_m)}.
\label{eq:Jij}
\eeq
When $\mu$ is larger than the transverse center-of-mass frequency, $J_{i,j}$ falls off with distance as roughly $J_{i,j} \sim J_0/|i-j|^\alpha$, where $J_0$ is of order $\approx 1$ kHz and $\alpha$ can be tuned between 0 and 3 using trap and laser parameters \cite{Porras2004,Islam2013}. 

The $V_{i,m}$ term in Eq. \ref{eqn:XYHam} is given by a similar formula,
\eq
V_{i,m} =  \frac{\left(\eta_{i,m} \Omega_i\right)^2}{8(\mu-\omega_m)}.
\eeq
For very long-ranged spin-spin interactions ($\alpha \lesssim 0.5$), or for small numbers of ions, the $V_{i,m}$ terms are approximately uniform across the spin chain. In these instances, the $V_{i,m}$ coefficient can be factored out of the sum over ions in Eq. \ref{eq:spin1XY}, leaving only global $S_z^i$ and $(S_z^i)^2$ terms. For shorter-range interactions or for longer chain lengths, the $V_{i,m}$ terms can be eliminated by adding an additional set of beat frequencies at $\omega_- - \mu$ and $\omega_+ + \mu$, which would generate Ising-type interactions between effective spin-1 particles using the M$\o$lmer-S$\o$rensen gate \cite{Molmer1999}.

\FigureTwo

The ions are initialized before each experiment by cooling the transverse modes near their ground state of motion ($\bar{n}\approx 0.05$) and optically pumping the spins to the $\ket{00\cdots}$ state. After applying the Hamiltonian in Eq. \ref{eq:spin1XY} for varying lengths of time, we measure the population of the state $\ket{0}$ at each site by imaging spin-dependent fluorescence \cite{Olmschenk2007} onto an intensified CCD camera and observing which ions are `dark'. Because both of the $\ket{\pm}$ states appear `bright' during the detection process and are scattered into an incoherent mixture of the $\ket{F=1}$ states, our current setup does not allow discrimination among all three possible spin states in a single experiment. However, we can measure the population of either $\ket{+}$ or $\ket{-}$ by repeating the experiment and applying a $\pi$ rotation to the appropriate $\ket{0}\leftrightarrow\ket{\pm}$ transition before the fluorescence imaging. For instance, measuring an ion in the `dark' state after a $\pi$ pulse between $\ket{0}\leftrightarrow\ket{+}$ indicates that the spin was in the $\ket{+}$ state before detection. This binary discrimination is not a fundamental limit to future experiments, since populations could be `shelved' into atomic states that do not participate in the detection cycle.

Since the ions are initialized to the $\ket{00\cdots}$ state, and because the spin-spin interactions in Eq. \ref{eq:spin1XY} conserve the quantity $\sum_i S_z^i \equiv \mathcal{S}_z$, the dynamics are restricted to the set of states with $\szsubspace$. The $\szsubspace$ subspace is protected against fluctuations in the real magnetic field $\Delta B(t)$, which would otherwise result in an unwanted noise term $\mu_B \Delta B(t) \mathcal{S}_z$ (where $\mu_B$ is the Bohr magneton). For instance, the $T_2$ coherence times of the $\ket{0}\leftrightarrow\ket{\pm}$ transitions were measured to be 0.5 ms, limited by magnetic field noise. Nevertheless, the data in Figures \ref{fig:dynamics2ions} and \ref{fig:ParityFig} (below) exhibit coherence and entanglement for several ms (limited by laser intensity noise), demonstrating the robustness of this `decoherence-free' subspace against time-varying magnetic fields. Remaining within this subspace does not substantially limit the size of the accessible Hilbert space, since the number of states in the $\szsubspace$ subspace of $N$ spin-1 particles scales as $\sim 3^N/(2\sqrt{N})$ for large $N$, which is exponentially greater than the $2^N$ states accessible in a spin-1/2 system.

\section{Coherent dynamics of two spins and entanglement verification}

For a system of 2 spins, dynamical evolution under the Hamiltonian in Eq. \ref{eq:spin1XY} can be understood as Rabi flopping between the $\ket{00}$ and $\left( \ket{+-} + \ket{-+} \right)/\sqrt{2}$ states with Rabi frequency $\sqrt{2} J_{1,2}$. This behavior is shown in Fig. \ref{fig:dynamics2ions}, where panels (a), (b), and (c) show the probability of each ion to be in the $\ket{0}$, $\ket{-}$, and $\ket{+}$ states, respectively. The population remains in the $\mathcal{S}_z=0$ subspace, as expected: Fig. \ref{fig:dynamics2ions}(a) shows the absence of the $\mathcal{S}_z \neq 0$ states ($\ket{0+}$, $\ket{0-}$, $\ket{+0}$, and, $\ket{-0}$), while Fig. \ref{fig:dynamics2ions}(b) and (c) respectively show the absence of the other $\mathcal{S}_z \neq 0$ states $\ket{--}$ and $\ket{++}$. The drift in $J_{1,2}$ evidenced in Fig. \ref{fig:dynamics2ions}(c) could be stabilized in future experiments by feeding back to the trap RF voltage to better stabilize the radial trap frequencies.

The different ions $i$ can experience position-dependent $S_z^i$ and $\left(S_z^i\right)^2$ shifts. We attribute this effect to a micromotion gradient, since the shifts can be compensated by adjustments of the voltages on the DC trap electrodes. The calculation overlaid in Fig. \ref{fig:dynamics2ions} includes the site-dependent terms $(200 $ Hz$) S_z^{(2)}$ + $(150$ Hz$) (S_z^{(2)})^2$, which were left as free fitting parameters when numerically evolving the Schr\"odinger equation under the Hamiltonian in Eq. \ref{eqn:XYHam}. The plotted curves assume strictly unitary evolution (i.e. no decoherence) over the timescale of the experiments.

\FigureThree

At a time $t=0.5/(\sqrt{2}J_{1,2})$, which is roughly 0.27 ms in Fig. \ref{fig:dynamics2ions}, the system is left approximately in the entangled state $\left( \ket{+-} + \ket{-+} \right)/\sqrt{2}$. To verify entanglement in the system, one could use spin-1 analogues of Bell-type inequalities \cite{Collins2002}, which require many local rotations but are sensitive to maximally entangled states like $\left(\ket{00}+\ket{+-}+\ket{-+}\right)/\sqrt{3}$. However, for the class of states generated by the XY interactions, a much simpler series of global rotations is sufficient to verify entanglement. The analysis consists of performing three sequential rotations on the $\ket{0}$ to $\ket{\pm}$ transitions, 
\eq
\label{eqn:Rotations}
R_{0\pm}(\theta,\phi) = e
^{\left( \frac{i\theta}{2} \sum_k [e^{\pm i\phi} (\ket{\pm}\!\bra{0})_k + e^{\mp i\phi} (\ket{0}\!\bra{\pm})_k] \right)},
\eeq 
before measuring the population in $\ket{0}$. The rotation sequence is given by $R_{0+}(\pi/2,\varphi)R_{0+}(\pi/2,0)R_{0-}(\pi,0)$, with the rotations applied from right to left. The first two rotations map the state $\left( \ket{+-} + \ket{-+} \right)/\sqrt{2}$ to $\left( \ket{00} + \ket{++} \right)/\sqrt{2}$, while the phase of the third rotation is varied to analyze the entanglement of this resulting state \cite{Sackett2000}. The parity $\Pi = \sum_{j=0}^2 (-1)^j P_j$ (with $P_j$ the probability of $j$ atoms in $\ket{0}$) oscillates as a function of the phase $\varphi$ of the third pulse, and the amplitude of its oscillation depends on the off-diagonal density matrix elements:
\small
\begin{eqnarray}
\Pi(\varphi) =& C + \frac{1}{2}\cos 2\varphi \left( P_{\mathsmaller{--}}+ P_{\mathsmaller{++}} - P_{\mathsmaller{+-}} - P_{\mathsmaller{-+}} \right. \\
&\left.  - 2 |\rho_\mathsmaller{+-,-+}| - 2 |\rho_\mathsmaller{--,++}| \right) \nonumber \\
+ \frac{1}{2} \sin& 2\varphi \left( 2|\rho_\mathsmaller{-+,++}| + 2|\rho_\mathsmaller{+-,++}| - 2|\rho_\mathsmaller{--,-+}| - 2|\rho_\mathsmaller{-+,--}| \right), \nonumber
\end{eqnarray}
\normalsize
where $P_{\mathsmaller{i}}$ is the population in state $\ket{i}$ ($\ket{i} = \ket{--},\ket{-+}$, etc.), $\rho_{i,j}$ is the off-diagonal density matrix element quantifying the coherence between $\ket{i}$ and $\ket{j}$, and $C$ is a constant offset that depends on the various density matrix elements but not on the phase $\varphi$ of the final rotation. The populations in $\ket{++}$ and $\ket{--}$ are negligible, simplifying this expression: 
\begin{equation}
\Pi(\varphi) \approx C - A\cos2\varphi
\end{equation}
where the oscillation amplitude
\eq
A = \frac{1}{2} \left(P_{\mathsmaller{+-}} + P_{\mathsmaller{-+}} + 2 |\rho_\mathsmaller{+-,-+}| \right)
\eeq 
is akin to the entanglement fidelity $\mathcal{F}$ of GHZ states in two-level systems \cite{Sackett2000}. Measuring the amplitude $A$ of the parity oscillation $\Pi(\varphi)$ then allows us to verify entanglement for certain classes of states. According to an analysis analogous to that in \cite{Sackett2000}, the following inequality holds for all separable qutrit states:
\eq
2A + P_{\mathsmaller{00}} + 2 |\rho_{\mathsmaller{+-,00}}| + 2 |\rho_{\mathsmaller{-+,00}}| \leq 1.
\label{eq:entanglementInequality}
\eeq
Hence, violation of this inequality demonstrates entanglement between spin-1 particles or qutrits, and measuring an amplitude of $A> 1/2$ is sufficient to violate the inequality.

Figure \ref{fig:ParityFig}(b) shows an example of the measured parity curve used to extract the amplitude $A$ and verify entanglement between the qutrit pair. Such measurements can be repeated for different durations of exposure to the XY Hamiltonian. At times $t = (2n+1)/(2\sqrt{2}J_{1,2})$ ($n=0,1,2,\ldots$), the system should again be in the state $\left( \ket{+-} + \ket{-+} \right)/\sqrt{2}$, while at times $t=n/(\sqrt{2}J_{1,2})$ it should return to the unentangled product state $\ket{00}$. The result is plotted in Fig. \ref{fig:ParityFig}(c). 

Two known sources of dephasing contribute to the observed loss of coherence in the experiment. First, laser intensity fluctuations and pointing instability cause noise in the spin-spin coupling term, leading to apparent dephasing when many repetitions are averaged together. These fluctuations could be compensated in future experiments by variants of the method of composite pulses \cite{Brown2004, Albrecht2014}. The second dephasing source results from inhomogeneities in the $V_{i,m}$ term (Eq. \ref{eqn:XYHam}) across the chain, which will cause different spins to acquire phases at different rates. This could be compensated by adding an extra driving term to cancel the inhomogeneities or by applying a series of echo pulses \cite{Cai2012,Biercuk2009PRA}. Fluctuating external magnetic fields and off-resonant coupling to the carrier transition would ordinarily add dephasing noise along the $\hat{z}$ direction, but have been suppressed here by working in the $\mathcal{S}_z=0$ subspace.

\FigureFour

\FigureFive

\section{Ground state production}

We can also add an effective $(S_z^i)^2$ field term, $D \sum_{i=1}^N (S_z^i)^2$, to the Hamiltonian by shifting the beat frequencies of the Raman lasers to $\omega_+ - \mu - D$ and $\omega_- + \mu - D$. This $(S_z^i)^2$ term can be used to adiabatically prepare the ground state of the XY Hamiltonian in Eq. \ref{eq:spin1XY}. As before, the spins are prepared in $\ket{00\cdots}$, which is the approximate ground state of Eq. \ref{eqn:XYHam} in the presence of a large (5 kHz) $(S_z^i)^2$ field. This field is then ramped down slowly according to $D(t) = (5 \:\mathrm{kHz}) e^{-t/(0.167 \:\mathrm{ms})}$. Figure \ref{fig:EvenIonGroundState} shows the populations measured at the end of the $(S_z^i)^2$ ramp for two and four spins, which match reasonably well with the calculated ground state. 

Measurements of populations in the $S_z$ basis necessarily discard phase information about components of the final state. This can be important in many spin models, including the XY model, where such measurements alone cannot discriminate between different eigenstates. For example, the ground state of an XY model with two spin-1 particles is $\ket{00}/\sqrt{2} - \left( \ket{-+} + \ket{+-} \right)/2$, while the highest excited state is $\ket{00}/\sqrt{2} + \left( \ket{-+} + \ket{+-} \right)/2$, differing only by a relative phase. We check that we are creating the ground state after our adiabatic protocol by applying a pair of rotations, $R_{0-}(\pi/2,\varphi)R_{0+}(\pi/2,0)$, and measuring the parity $\Pi$ as was done in the entanglement analysis. This is expected to result in $\Pi(\varphi) = \frac{3}{8} \pm \frac{1}{2}\cos\varphi$, where the + and - correspond to the ground and highest excited states, respectively. As shown in Fig. \ref{fig:GroundStatePhase}, our measurements are consistent with having prepared the 2-spin ground state.

\section{Toward Haldane physics}
A more long-term goal for spin-1 quantum simulations will be to produce and study ground states in the Haldane phase \cite{Cohen2014}. It is known that an XY model with both nearest-neighbor and next-nearest-neighbor interactions can exhibit a symmetry-protected Haldane phase \cite{Murashima2005}, and it remains an open question whether a generic long-range XY model would show the same behavior. Already with our experimentally implemented Hamiltonian, we find a useful test case where the symmetry of the ground state prevents it from being created via the simple adiabatic protocol described above. 

The ground state $\ket{\psi}_{gs}$ of a long-range XY model can be calculated exactly for three spins. For our experimental coupling strengths $J_{i,j} \sim 1/|i-j|^{0.36}$,
\eqarray
\nonumber 
\ket{\psi}_{gs} = &\sqrt{0.16}& \left(\ket{0-+} -\ket{0+-} + \ket{-+0} - \ket{+-0} \right) \\ 
+& \sqrt{0.18}& \left(\ket{+0-} - \ket{-0+}\right). \label{eqn:XYgroundstate}
\eeqarray
This state has a 99.9\% overlap with a three-spin AKLT state \cite{Affleck1988}, which is the canonical example of a ground state in the Haldane phase that can be written in closed form for any number of spins. The state in Eq. \ref{eqn:XYgroundstate} is antisymmetric with respect to the same symmetries that govern the Haldane phase, such as left-right spatial inversion of the chain or a global rotation about $S_x$ by $\pi$ (which sends $\ket{+}$ to $\ket{-}$ and vice versa). However, since the starting state $\ket{000}$ and the applied Hamiltonian are symmetric with respect to these operations, we should be unable to reach the antisymmetric ground state with a simple adiabatic ramp. Indeed, we find numerically that a first order phase transition separates the symmetric and antisymmetric ground states, which cannot be adiabatically connected without breaking inversion and rotational symmetry. For the three-spin experiment in Fig. \ref{fig:3ionGroundState}, we hence prepare a state close to the first excited state rather than the ground state. This observation suggests that even in the presence of various experimental imperfections, the ground state of our three-spin XY model enjoys the same symmetry protection as the Haldane phase. 

\FigureSix

In this paper, we have demonstrated the basic ingredients which are needed for the implementation of quantum simulations with spins greater than 1/2. We believe that this work opens paths for studying the exciting physics beyond spin-1/2 systems, and we have already taken the first steps towards exploring the richness of topological phases. In particular, for a long-range spin-1 XY model, we have demonstrated coherent Schr\"odinger evolution and the capability to create symmetric ground states. We have observed that for odd numbers of spins, symmetry considerations prevent us from creating ground states which bear a close resemblance to AKLT states and hence may belong to the Haldane phase. Future work will address the questions of how to add a Heisenberg term and symmetry-breaking perturbations to the Hamiltonian so as to prepare antisymmetric ground states \cite{Cohen2014}, which will allow us to create and probe interesting edge states in the Haldane phase.

\section{Acknowledgments}
We thank Brian Neyenhuis, Paul Hess, Alexey Gorshkov, and Zhe-Xuan Gong for critical discussions. This work is supported by the U.S. Army Research Office (ARO) Award W911NF0710576 with funds from the DARPA Optical Lattice Emulator Program, ARO award W911NF0410234 with funds from the IARPA MQCO Program, and the NSF Physics Frontier Center at JQI.

\bibliography{../Bibliography/lotsofrefs}

\begin{thebibliography}{46}%
\makeatletter
\providecommand \@ifxundefined [1]{%
 \@ifx{#1\undefined}
}%
\providecommand \@ifnum [1]{%
 \ifnum #1\expandafter \@firstoftwo
 \else \expandafter \@secondoftwo
 \fi
}%
\providecommand \@ifx [1]{%
 \ifx #1\expandafter \@firstoftwo
 \else \expandafter \@secondoftwo
 \fi
}%
\providecommand \natexlab [1]{#1}%
\providecommand \enquote  [1]{``#1''}%
\providecommand \bibnamefont  [1]{#1}%
\providecommand \bibfnamefont [1]{#1}%
\providecommand \citenamefont [1]{#1}%
\providecommand \href@noop [0]{\@secondoftwo}%
\providecommand \href [0]{\begingroup \@sanitize@url \@href}%
\providecommand \@href[1]{\@@startlink{#1}\@@href}%
\providecommand \@@href[1]{\endgroup#1\@@endlink}%
\providecommand \@sanitize@url [0]{\catcode `\\12\catcode `\$12\catcode
  `\&12\catcode `\#12\catcode `\^12\catcode `\_12\catcode `\%12\relax}%
\providecommand \@@startlink[1]{}%
\providecommand \@@endlink[0]{}%
\providecommand \url  [0]{\begingroup\@sanitize@url \@url }%
\providecommand \@url [1]{\endgroup\@href {#1}{\urlprefix }}%
\providecommand \urlprefix  [0]{URL }%
\providecommand \Eprint [0]{\href }%
\providecommand \doibase [0]{http://dx.doi.org/}%
\providecommand \selectlanguage [0]{\@gobble}%
\providecommand \bibinfo  [0]{\@secondoftwo}%
\providecommand \bibfield  [0]{\@secondoftwo}%
\providecommand \translation [1]{[#1]}%
\providecommand \BibitemOpen [0]{}%
\providecommand \bibitemStop [0]{}%
\providecommand \bibitemNoStop [0]{.\EOS\space}%
\providecommand \EOS [0]{\spacefactor3000\relax}%
\providecommand \BibitemShut  [1]{\csname bibitem#1\endcsname}%
\let\auto@bib@innerbib\@empty
\bibitem [{\citenamefont {Scarani}\ \emph {et~al.}(2009)\citenamefont
  {Scarani}, \citenamefont {Bechmann-Pasquinucci}, \citenamefont {Cerf},
  \citenamefont {Dusek}, \citenamefont {Lutkenhaus},\ and\ \citenamefont
  {Peev}}]{Scarani2009}%
  \BibitemOpen
  \bibfield  {author} {\bibinfo {author} {\bibfnamefont {V.}~\bibnamefont
  {Scarani}}, \bibinfo {author} {\bibfnamefont {H.}~\bibnamefont
  {Bechmann-Pasquinucci}}, \bibinfo {author} {\bibfnamefont {N.~J.}\
  \bibnamefont {Cerf}}, \bibinfo {author} {\bibfnamefont {M.}~\bibnamefont
  {Dusek}}, \bibinfo {author} {\bibfnamefont {N.}~\bibnamefont {Lutkenhaus}}, \
  and\ \bibinfo {author} {\bibfnamefont {M.}~\bibnamefont {Peev}},\ }\href@noop
  {} {\bibfield  {journal} {\bibinfo  {journal} {Rev. Mod. Phys.}\ }\textbf
  {\bibinfo {volume} {81}},\ \bibinfo {pages} {1301} (\bibinfo {year}
  {2009})}\BibitemShut {NoStop}%
\bibitem [{\citenamefont {Ladd}\ \emph {et~al.}(2010)\citenamefont {Ladd},
  \citenamefont {Jelezko}, \citenamefont {Laflamme}, \citenamefont {Nakamura},
  \citenamefont {Monroe},\ and\ \citenamefont {O'Brien}}]{Ladd2010}%
  \BibitemOpen
  \bibfield  {author} {\bibinfo {author} {\bibfnamefont {T.~D.}\ \bibnamefont
  {Ladd}}, \bibinfo {author} {\bibfnamefont {F.}~\bibnamefont {Jelezko}},
  \bibinfo {author} {\bibfnamefont {R.}~\bibnamefont {Laflamme}}, \bibinfo
  {author} {\bibfnamefont {Y.}~\bibnamefont {Nakamura}}, \bibinfo {author}
  {\bibfnamefont {C.}~\bibnamefont {Monroe}}, \ and\ \bibinfo {author}
  {\bibfnamefont {J.~L.}\ \bibnamefont {O'Brien}},\ }\href@noop {} {\bibfield
  {journal} {\bibinfo  {journal} {Nature}\ }\textbf {\bibinfo {volume} {464}},\
  \bibinfo {pages} {45} (\bibinfo {year} {2010})}\BibitemShut {NoStop}%
\bibitem [{\citenamefont {Bloch}\ \emph {et~al.}(2012)\citenamefont {Bloch},
  \citenamefont {Dalibard},\ and\ \citenamefont {Nascimbene}}]{Bloch2012}%
  \BibitemOpen
  \bibfield  {author} {\bibinfo {author} {\bibfnamefont {I.}~\bibnamefont
  {Bloch}}, \bibinfo {author} {\bibfnamefont {J.}~\bibnamefont {Dalibard}}, \
  and\ \bibinfo {author} {\bibfnamefont {S.}~\bibnamefont {Nascimbene}},\
  }\href@noop {} {\bibfield  {journal} {\bibinfo  {journal} {Nature Physics}\
  }\textbf {\bibinfo {volume} {8}},\ \bibinfo {pages} {267} (\bibinfo {year}
  {2012})}\BibitemShut {NoStop}%
\bibitem [{\citenamefont {Blatt}\ and\ \citenamefont {Roos}(2012)}]{Blatt2012}%
  \BibitemOpen
  \bibfield  {author} {\bibinfo {author} {\bibfnamefont {R.}~\bibnamefont
  {Blatt}}\ and\ \bibinfo {author} {\bibfnamefont {C.~F.}\ \bibnamefont
  {Roos}},\ }\href@noop {} {\bibfield  {journal} {\bibinfo  {journal} {Nature
  Physics}\ }\textbf {\bibinfo {volume} {8}},\ \bibinfo {pages} {277} (\bibinfo
  {year} {2012})}\BibitemShut {NoStop}%
\bibitem [{\citenamefont {Auerbach}(1994)}]{Auerbach1994}%
  \BibitemOpen
  \bibfield  {author} {\bibinfo {author} {\bibfnamefont {A.}~\bibnamefont
  {Auerbach}},\ }\href@noop {} {\emph {\bibinfo {title} {Interacting Electrons
  and Quantum Magnetism}}}\ (\bibinfo  {publisher} {Springer},\ \bibinfo
  {address} {New York},\ \bibinfo {year} {1994})\BibitemShut {NoStop}%
\bibitem [{\citenamefont {Aharonov}\ \emph {et~al.}(2007)\citenamefont
  {Aharonov}, \citenamefont {Gottesman}, \citenamefont {Irani},\ and\
  \citenamefont {Kempe}}]{Aharonov2007}%
  \BibitemOpen
  \bibfield  {author} {\bibinfo {author} {\bibfnamefont {D.}~\bibnamefont
  {Aharonov}}, \bibinfo {author} {\bibfnamefont {D.}~\bibnamefont {Gottesman}},
  \bibinfo {author} {\bibfnamefont {S.}~\bibnamefont {Irani}}, \ and\ \bibinfo
  {author} {\bibfnamefont {J.}~\bibnamefont {Kempe}},\ }in\ \href {\doibase
  10.1109/FOCS.2007.46} {\emph {\bibinfo {booktitle} {Foundations of Computer
  Science, 2007. FOCS '07. 48th Annual IEEE Symposium on}}}\ (\bibinfo {year}
  {2007})\ pp.\ \bibinfo {pages} {373--383}\BibitemShut {NoStop}%
\bibitem [{\citenamefont {Hallgren}\ \emph {et~al.}()\citenamefont {Hallgren},
  \citenamefont {Nagaj},\ and\ \citenamefont {Narayanaswami}}]{HallgrenArxiv}%
  \BibitemOpen
  \bibfield  {author} {\bibinfo {author} {\bibfnamefont {S.}~\bibnamefont
  {Hallgren}}, \bibinfo {author} {\bibfnamefont {D.}~\bibnamefont {Nagaj}}, \
  and\ \bibinfo {author} {\bibfnamefont {S.}~\bibnamefont {Narayanaswami}},\
  }\href@noop {} {\enquote {\bibinfo {title} {The local hamiltonian problem on
  a line with eight states is qma-complete},}\ }\bibinfo {howpublished}
  {arXiv:1312.1469}\BibitemShut {NoStop}%
\bibitem [{\citenamefont {Lanyon}\ \emph {et~al.}(2009)\citenamefont {Lanyon},
  \citenamefont {Barbieri}, \citenamefont {Almeida}, \citenamefont {Jennewein},
  \citenamefont {Ralph}, \citenamefont {Resch}, \citenamefont {Pryde},
  \citenamefont {O'Brien}, \citenamefont {Gilchrist},\ and\ \citenamefont
  {White}}]{Lanyon2009}%
  \BibitemOpen
  \bibfield  {author} {\bibinfo {author} {\bibfnamefont {B.~P.}\ \bibnamefont
  {Lanyon}}, \bibinfo {author} {\bibfnamefont {M.}~\bibnamefont {Barbieri}},
  \bibinfo {author} {\bibfnamefont {M.~P.}\ \bibnamefont {Almeida}}, \bibinfo
  {author} {\bibfnamefont {T.}~\bibnamefont {Jennewein}}, \bibinfo {author}
  {\bibfnamefont {T.~C.}\ \bibnamefont {Ralph}}, \bibinfo {author}
  {\bibfnamefont {K.~J.}\ \bibnamefont {Resch}}, \bibinfo {author}
  {\bibfnamefont {G.~J.}\ \bibnamefont {Pryde}}, \bibinfo {author}
  {\bibfnamefont {J.~L.}\ \bibnamefont {O'Brien}}, \bibinfo {author}
  {\bibfnamefont {A.}~\bibnamefont {Gilchrist}}, \ and\ \bibinfo {author}
  {\bibfnamefont {A.~G.}\ \bibnamefont {White}},\ }\href@noop {} {\bibfield
  {journal} {\bibinfo  {journal} {Nature Physics}\ }\textbf {\bibinfo {volume}
  {5}},\ \bibinfo {pages} {134} (\bibinfo {year} {2009})}\BibitemShut {NoStop}%
\bibitem [{\citenamefont {Brukner}\ \emph {et~al.}(2002)\citenamefont
  {Brukner}, \citenamefont {Zukowski},\ and\ \citenamefont
  {Zeilinger}}]{Brukner2002}%
  \BibitemOpen
  \bibfield  {author} {\bibinfo {author} {\bibfnamefont {C.}~\bibnamefont
  {Brukner}}, \bibinfo {author} {\bibfnamefont {M.}~\bibnamefont {Zukowski}}, \
  and\ \bibinfo {author} {\bibfnamefont {A.}~\bibnamefont {Zeilinger}},\
  }\href@noop {} {\bibfield  {journal} {\bibinfo  {journal} {Phys. Rev. Lett.}\
  }\textbf {\bibinfo {volume} {89}},\ \bibinfo {pages} {197901} (\bibinfo
  {year} {2002})}\BibitemShut {NoStop}%
\bibitem [{\citenamefont {Haldane}(1983{\natexlab{a}})}]{Haldane1983}%
  \BibitemOpen
  \bibfield  {author} {\bibinfo {author} {\bibfnamefont {F.~D.~M.}\
  \bibnamefont {Haldane}},\ }\href@noop {} {\bibfield  {journal} {\bibinfo
  {journal} {Physics Letters A}\ }\textbf {\bibinfo {volume} {93}},\ \bibinfo
  {pages} {464} (\bibinfo {year} {1983}{\natexlab{a}})}\BibitemShut {NoStop}%
\bibitem [{\citenamefont {Haldane}(1983{\natexlab{b}})}]{Haldane1983a}%
  \BibitemOpen
  \bibfield  {author} {\bibinfo {author} {\bibfnamefont {F.~D.~M.}\
  \bibnamefont {Haldane}},\ }\href@noop {} {\bibfield  {journal} {\bibinfo
  {journal} {Phys. Rev. Lett.}\ }\textbf {\bibinfo {volume} {50}},\ \bibinfo
  {pages} {1153} (\bibinfo {year} {1983}{\natexlab{b}})}\BibitemShut {NoStop}%
\bibitem [{\citenamefont {Pollmann}\ \emph {et~al.}(2010)\citenamefont
  {Pollmann}, \citenamefont {Turner}, \citenamefont {Berg},\ and\ \citenamefont
  {Oshikawa}}]{Pollmann2010}%
  \BibitemOpen
  \bibfield  {author} {\bibinfo {author} {\bibfnamefont {F.}~\bibnamefont
  {Pollmann}}, \bibinfo {author} {\bibfnamefont {A.~M.}\ \bibnamefont
  {Turner}}, \bibinfo {author} {\bibfnamefont {E.}~\bibnamefont {Berg}}, \ and\
  \bibinfo {author} {\bibfnamefont {M.}~\bibnamefont {Oshikawa}},\ }\href@noop
  {} {\bibfield  {journal} {\bibinfo  {journal} {Phys. Rev. B}\ }\textbf
  {\bibinfo {volume} {81}},\ \bibinfo {pages} {064439} (\bibinfo {year}
  {2010})}\BibitemShut {NoStop}%
\bibitem [{\citenamefont {Kennedy}\ and\ \citenamefont
  {Tasaki}(1992{\natexlab{a}})}]{Kennedy1992}%
  \BibitemOpen
  \bibfield  {author} {\bibinfo {author} {\bibfnamefont {T.}~\bibnamefont
  {Kennedy}}\ and\ \bibinfo {author} {\bibfnamefont {H.}~\bibnamefont
  {Tasaki}},\ }\href@noop {} {\bibfield  {journal} {\bibinfo  {journal} {Phys.
  Rev. B}\ }\textbf {\bibinfo {volume} {45}},\ \bibinfo {pages} {304} (\bibinfo
  {year} {1992}{\natexlab{a}})}\BibitemShut {NoStop}%
\bibitem [{\citenamefont {Kennedy}\ and\ \citenamefont
  {Tasaki}(1992{\natexlab{b}})}]{Kennedy1992a}%
  \BibitemOpen
  \bibfield  {author} {\bibinfo {author} {\bibfnamefont {T.}~\bibnamefont
  {Kennedy}}\ and\ \bibinfo {author} {\bibfnamefont {H.}~\bibnamefont
  {Tasaki}},\ }\href@noop {} {\bibfield  {journal} {\bibinfo  {journal}
  {Commun. Math. Phys.}\ }\textbf {\bibinfo {volume} {147}},\ \bibinfo {pages}
  {431} (\bibinfo {year} {1992}{\natexlab{b}})}\BibitemShut {NoStop}%
\bibitem [{\citenamefont {Balents}(2010)}]{Balents2010}%
  \BibitemOpen
  \bibfield  {author} {\bibinfo {author} {\bibfnamefont {L.}~\bibnamefont
  {Balents}},\ }\href@noop {} {\bibfield  {journal} {\bibinfo  {journal}
  {Nature}\ }\textbf {\bibinfo {volume} {464}},\ \bibinfo {pages} {199}
  (\bibinfo {year} {2010})}\BibitemShut {NoStop}%
\bibitem [{\citenamefont {Darmawan}\ and\ \citenamefont
  {Bartlett}(2010)}]{Darmawan2010}%
  \BibitemOpen
  \bibfield  {author} {\bibinfo {author} {\bibfnamefont {A.~S.}\ \bibnamefont
  {Darmawan}}\ and\ \bibinfo {author} {\bibfnamefont {S.~D.}\ \bibnamefont
  {Bartlett}},\ }\href@noop {} {\bibfield  {journal} {\bibinfo  {journal}
  {Phys. Rev. A}\ }\textbf {\bibinfo {volume} {82}},\ \bibinfo {pages} {012328}
  (\bibinfo {year} {2010})}\BibitemShut {NoStop}%
\bibitem [{\citenamefont {Asoudeh}\ and\ \citenamefont
  {Karimipour}(2013)}]{Asoudeh2013}%
  \BibitemOpen
  \bibfield  {author} {\bibinfo {author} {\bibfnamefont {M.}~\bibnamefont
  {Asoudeh}}\ and\ \bibinfo {author} {\bibfnamefont {V.}~\bibnamefont
  {Karimipour}},\ }\href@noop {} {\bibfield  {journal} {\bibinfo  {journal}
  {Quantum Information Processing}\ }\textbf {\bibinfo {volume} {13}},\
  \bibinfo {pages} {601} (\bibinfo {year} {2013})}\BibitemShut {NoStop}%
\bibitem [{\citenamefont {Else}\ \emph
  {et~al.}(2012{\natexlab{a}})\citenamefont {Else}, \citenamefont {Bartlett},\
  and\ \citenamefont {Doherty}}]{Else2012NJP}%
  \BibitemOpen
  \bibfield  {author} {\bibinfo {author} {\bibfnamefont {D.~V.}\ \bibnamefont
  {Else}}, \bibinfo {author} {\bibfnamefont {S.~D.}\ \bibnamefont {Bartlett}},
  \ and\ \bibinfo {author} {\bibfnamefont {A.~C.}\ \bibnamefont {Doherty}},\
  }\href@noop {} {\bibfield  {journal} {\bibinfo  {journal} {New Journal of
  Physics}\ }\textbf {\bibinfo {volume} {14}},\ \bibinfo {pages} {113016}
  (\bibinfo {year} {2012}{\natexlab{a}})}\BibitemShut {NoStop}%
\bibitem [{\citenamefont {Brennen}\ and\ \citenamefont
  {Miyake}(2008)}]{Brennen2008}%
  \BibitemOpen
  \bibfield  {author} {\bibinfo {author} {\bibfnamefont {G.~K.}\ \bibnamefont
  {Brennen}}\ and\ \bibinfo {author} {\bibfnamefont {A.}~\bibnamefont
  {Miyake}},\ }\href@noop {} {\bibfield  {journal} {\bibinfo  {journal} {Phys.
  Rev. Lett.}\ }\textbf {\bibinfo {volume} {101}},\ \bibinfo {pages} {010502}
  (\bibinfo {year} {2008})}\BibitemShut {NoStop}%
\bibitem [{\citenamefont {Else}\ \emph
  {et~al.}(2012{\natexlab{b}})\citenamefont {Else}, \citenamefont {Schwarz},
  \citenamefont {Bartlett},\ and\ \citenamefont {Doherty}}]{Else2012PRL}%
  \BibitemOpen
  \bibfield  {author} {\bibinfo {author} {\bibfnamefont {D.~V.}\ \bibnamefont
  {Else}}, \bibinfo {author} {\bibfnamefont {I.}~\bibnamefont {Schwarz}},
  \bibinfo {author} {\bibfnamefont {S.~D.}\ \bibnamefont {Bartlett}}, \ and\
  \bibinfo {author} {\bibfnamefont {A.~C.}\ \bibnamefont {Doherty}},\
  }\href@noop {} {\bibfield  {journal} {\bibinfo  {journal} {Phys. Rev. Lett.}\
  }\textbf {\bibinfo {volume} {108}},\ \bibinfo {pages} {240505} (\bibinfo
  {year} {2012}{\natexlab{b}})}\BibitemShut {NoStop}%
\bibitem [{\citenamefont {Lanyon}\ \emph {et~al.}(2008)\citenamefont {Lanyon},
  \citenamefont {Weinhold}, \citenamefont {Langford}, \citenamefont {O'Brien},
  \citenamefont {Resch}, \citenamefont {Gilchrist},\ and\ \citenamefont
  {White}}]{Lanyon2008}%
  \BibitemOpen
  \bibfield  {author} {\bibinfo {author} {\bibfnamefont {B.~P.}\ \bibnamefont
  {Lanyon}}, \bibinfo {author} {\bibfnamefont {T.~J.}\ \bibnamefont
  {Weinhold}}, \bibinfo {author} {\bibfnamefont {N.~K.}\ \bibnamefont
  {Langford}}, \bibinfo {author} {\bibfnamefont {J.~L.}\ \bibnamefont
  {O'Brien}}, \bibinfo {author} {\bibfnamefont {K.~J.}\ \bibnamefont {Resch}},
  \bibinfo {author} {\bibfnamefont {A.}~\bibnamefont {Gilchrist}}, \ and\
  \bibinfo {author} {\bibfnamefont {A.~G.}\ \bibnamefont {White}},\ }\href@noop
  {} {\bibfield  {journal} {\bibinfo  {journal} {Phys. Rev. Lett.}\ }\textbf
  {\bibinfo {volume} {100}},\ \bibinfo {pages} {060504} (\bibinfo {year}
  {2008})}\BibitemShut {NoStop}%
\bibitem [{\citenamefont {Bianchetti}\ \emph {et~al.}(2010)\citenamefont
  {Bianchetti}, \citenamefont {Filipp}, \citenamefont {Baur}, \citenamefont
  {Fink}, \citenamefont {Lang}, \citenamefont {Steffen}, \citenamefont
  {Boissonneault}, \citenamefont {Blais},\ and\ \citenamefont
  {Wallraff}}]{Bianchetti2010}%
  \BibitemOpen
  \bibfield  {author} {\bibinfo {author} {\bibfnamefont {R.}~\bibnamefont
  {Bianchetti}}, \bibinfo {author} {\bibfnamefont {S.}~\bibnamefont {Filipp}},
  \bibinfo {author} {\bibfnamefont {M.}~\bibnamefont {Baur}}, \bibinfo {author}
  {\bibfnamefont {J.~M.}\ \bibnamefont {Fink}}, \bibinfo {author}
  {\bibfnamefont {C.}~\bibnamefont {Lang}}, \bibinfo {author} {\bibfnamefont
  {L.}~\bibnamefont {Steffen}}, \bibinfo {author} {\bibfnamefont
  {M.}~\bibnamefont {Boissonneault}}, \bibinfo {author} {\bibfnamefont
  {A.}~\bibnamefont {Blais}}, \ and\ \bibinfo {author} {\bibfnamefont
  {A.}~\bibnamefont {Wallraff}},\ }\href@noop {} {\bibfield  {journal}
  {\bibinfo  {journal} {Phys. Rev. Lett.}\ }\textbf {\bibinfo {volume} {105}},\
  \bibinfo {pages} {223601} (\bibinfo {year} {2010})}\BibitemShut {NoStop}%
\bibitem [{\citenamefont {Chang}\ \emph {et~al.}(2005)\citenamefont {Chang},
  \citenamefont {Qin}, \citenamefont {Zhang}, \citenamefont {You},\ and\
  \citenamefont {Chapman}}]{Chang2005}%
  \BibitemOpen
  \bibfield  {author} {\bibinfo {author} {\bibfnamefont {M.-S.}\ \bibnamefont
  {Chang}}, \bibinfo {author} {\bibfnamefont {Q.}~\bibnamefont {Qin}}, \bibinfo
  {author} {\bibfnamefont {W.}~\bibnamefont {Zhang}}, \bibinfo {author}
  {\bibfnamefont {L.}~\bibnamefont {You}}, \ and\ \bibinfo {author}
  {\bibfnamefont {M.~S.}\ \bibnamefont {Chapman}},\ }\href@noop {} {\bibfield
  {journal} {\bibinfo  {journal} {Nat. Phys.}\ }\textbf {\bibinfo {volume}
  {1}},\ \bibinfo {pages} {111} (\bibinfo {year} {2005})}\BibitemShut {NoStop}%
\bibitem [{\citenamefont {Stamper-Kurn}\ and\ \citenamefont
  {Ueda}(2013)}]{Stamper-Kurn2013}%
  \BibitemOpen
  \bibfield  {author} {\bibinfo {author} {\bibfnamefont {D.~M.}\ \bibnamefont
  {Stamper-Kurn}}\ and\ \bibinfo {author} {\bibfnamefont {M.}~\bibnamefont
  {Ueda}},\ }\href@noop {} {\bibfield  {journal} {\bibinfo  {journal} {Rev.
  Mod. Phys.}\ }\textbf {\bibinfo {volume} {85}},\ \bibinfo {pages} {1191}
  (\bibinfo {year} {2013})}\BibitemShut {NoStop}%
\bibitem [{\citenamefont {Parker}\ \emph {et~al.}(2013)\citenamefont {Parker},
  \citenamefont {Ha},\ and\ \citenamefont {Chin}}]{Parker2013}%
  \BibitemOpen
  \bibfield  {author} {\bibinfo {author} {\bibfnamefont {C.~V.}\ \bibnamefont
  {Parker}}, \bibinfo {author} {\bibfnamefont {L.-C.}\ \bibnamefont {Ha}}, \
  and\ \bibinfo {author} {\bibfnamefont {C.}~\bibnamefont {Chin}},\ }\href@noop
  {} {\bibfield  {journal} {\bibinfo  {journal} {Nature Physics}\ }\textbf
  {\bibinfo {volume} {9}},\ \bibinfo {pages} {769} (\bibinfo {year}
  {2013})}\BibitemShut {NoStop}%
\bibitem [{\citenamefont {Cohen}\ and\ \citenamefont
  {Retzker}(2014)}]{Cohen2014}%
  \BibitemOpen
  \bibfield  {author} {\bibinfo {author} {\bibfnamefont {I.}~\bibnamefont
  {Cohen}}\ and\ \bibinfo {author} {\bibfnamefont {A.}~\bibnamefont
  {Retzker}},\ }\href@noop {} {\bibfield  {journal} {\bibinfo  {journal} {Phys.
  Rev. Lett.}\ }\textbf {\bibinfo {volume} {112}},\ \bibinfo {pages} {040503}
  (\bibinfo {year} {2014})}\BibitemShut {NoStop}%
\bibitem [{\citenamefont {Richerme}\ \emph {et~al.}(2014)\citenamefont
  {Richerme}, \citenamefont {Gong}, \citenamefont {Lee}, \citenamefont {Senko},
  \citenamefont {Smith}, \citenamefont {Foss-Feig}, \citenamefont {Michalakis},
  \citenamefont {Gorshkov},\ and\ \citenamefont {Monroe}}]{Richerme2014}%
  \BibitemOpen
  \bibfield  {author} {\bibinfo {author} {\bibfnamefont {P.}~\bibnamefont
  {Richerme}}, \bibinfo {author} {\bibfnamefont {Z.-X.}\ \bibnamefont {Gong}},
  \bibinfo {author} {\bibfnamefont {A.}~\bibnamefont {Lee}}, \bibinfo {author}
  {\bibfnamefont {C.}~\bibnamefont {Senko}}, \bibinfo {author} {\bibfnamefont
  {J.}~\bibnamefont {Smith}}, \bibinfo {author} {\bibfnamefont
  {M.}~\bibnamefont {Foss-Feig}}, \bibinfo {author} {\bibfnamefont
  {S.}~\bibnamefont {Michalakis}}, \bibinfo {author} {\bibfnamefont {A.~V.}\
  \bibnamefont {Gorshkov}}, \ and\ \bibinfo {author} {\bibfnamefont
  {C.}~\bibnamefont {Monroe}},\ }\href@noop {} {\bibfield  {journal} {\bibinfo
  {journal} {Nature}\ }\textbf {\bibinfo {volume} {511}},\ \bibinfo {pages}
  {198} (\bibinfo {year} {2014})}\BibitemShut {NoStop}%
\bibitem [{\citenamefont {Kim}\ \emph {et~al.}(2010)\citenamefont {Kim},
  \citenamefont {Chang}, \citenamefont {Korenblit}, \citenamefont {Islam},
  \citenamefont {Edwards}, \citenamefont {Freericks}, \citenamefont {Lin},
  \citenamefont {Duan},\ and\ \citenamefont {Monroe}}]{Kim2010}%
  \BibitemOpen
  \bibfield  {author} {\bibinfo {author} {\bibfnamefont {K.}~\bibnamefont
  {Kim}}, \bibinfo {author} {\bibfnamefont {M.-S.}\ \bibnamefont {Chang}},
  \bibinfo {author} {\bibfnamefont {S.}~\bibnamefont {Korenblit}}, \bibinfo
  {author} {\bibfnamefont {R.}~\bibnamefont {Islam}}, \bibinfo {author}
  {\bibfnamefont {E.~E.}\ \bibnamefont {Edwards}}, \bibinfo {author}
  {\bibfnamefont {J.~K.}\ \bibnamefont {Freericks}}, \bibinfo {author}
  {\bibfnamefont {G.-D.}\ \bibnamefont {Lin}}, \bibinfo {author} {\bibfnamefont
  {L.-M.}\ \bibnamefont {Duan}}, \ and\ \bibinfo {author} {\bibfnamefont
  {C.}~\bibnamefont {Monroe}},\ }\href@noop {} {\bibfield  {journal} {\bibinfo
  {journal} {Nature}\ }\textbf {\bibinfo {volume} {465}},\ \bibinfo {pages}
  {590} (\bibinfo {year} {2010})}\BibitemShut {NoStop}%
\bibitem [{\citenamefont {Richerme}\ \emph {et~al.}(2013)\citenamefont
  {Richerme}, \citenamefont {Senko}, \citenamefont {Smith}, \citenamefont
  {Lee}, \citenamefont {Korenblit},\ and\ \citenamefont
  {Monroe}}]{Richerme2013}%
  \BibitemOpen
  \bibfield  {author} {\bibinfo {author} {\bibfnamefont {P.}~\bibnamefont
  {Richerme}}, \bibinfo {author} {\bibfnamefont {C.}~\bibnamefont {Senko}},
  \bibinfo {author} {\bibfnamefont {J.}~\bibnamefont {Smith}}, \bibinfo
  {author} {\bibfnamefont {A.}~\bibnamefont {Lee}}, \bibinfo {author}
  {\bibfnamefont {S.}~\bibnamefont {Korenblit}}, \ and\ \bibinfo {author}
  {\bibfnamefont {C.}~\bibnamefont {Monroe}},\ }\href@noop {} {\bibfield
  {journal} {\bibinfo  {journal} {Phys. Rev. A}\ }\textbf {\bibinfo {volume}
  {88}},\ \bibinfo {pages} {012334} (\bibinfo {year} {2013})}\BibitemShut
  {NoStop}%
\bibitem [{\citenamefont {Lidar}\ \emph {et~al.}(1998)\citenamefont {Lidar},
  \citenamefont {Chuang},\ and\ \citenamefont {Whaley}}]{Lidar1998}%
  \BibitemOpen
  \bibfield  {author} {\bibinfo {author} {\bibfnamefont {D.~A.}\ \bibnamefont
  {Lidar}}, \bibinfo {author} {\bibfnamefont {I.~L.}\ \bibnamefont {Chuang}}, \
  and\ \bibinfo {author} {\bibfnamefont {K.~B.}\ \bibnamefont {Whaley}},\
  }\href@noop {} {\bibfield  {journal} {\bibinfo  {journal} {Phys. Rev. Lett.}\
  }\textbf {\bibinfo {volume} {81}},\ \bibinfo {pages} {2594} (\bibinfo {year}
  {1998})}\BibitemShut {NoStop}%
\bibitem [{\citenamefont {Kielpinski}\ \emph {et~al.}(2001)\citenamefont
  {Kielpinski}, \citenamefont {Meyer}, \citenamefont {Rowe}, \citenamefont
  {Sackett}, \citenamefont {Itano}, \citenamefont {Monroe},\ and\ \citenamefont
  {Wineland}}]{Kielpinski2001}%
  \BibitemOpen
  \bibfield  {author} {\bibinfo {author} {\bibfnamefont {D.}~\bibnamefont
  {Kielpinski}}, \bibinfo {author} {\bibfnamefont {V.}~\bibnamefont {Meyer}},
  \bibinfo {author} {\bibfnamefont {M.~A.}\ \bibnamefont {Rowe}}, \bibinfo
  {author} {\bibfnamefont {C.~A.}\ \bibnamefont {Sackett}}, \bibinfo {author}
  {\bibfnamefont {W.~M.}\ \bibnamefont {Itano}}, \bibinfo {author}
  {\bibfnamefont {C.}~\bibnamefont {Monroe}}, \ and\ \bibinfo {author}
  {\bibfnamefont {D.~J.}\ \bibnamefont {Wineland}},\ }\href@noop {} {\bibfield
  {journal} {\bibinfo  {journal} {Science}\ }\textbf {\bibinfo {volume}
  {291}},\ \bibinfo {pages} {1013} (\bibinfo {year} {2001})}\BibitemShut
  {NoStop}%
\bibitem [{\citenamefont {Grass}\ \emph {et~al.}(2013)\citenamefont {Grass},
  \citenamefont {Julia-Diaz}, \citenamefont {Kus},\ and\ \citenamefont
  {Lewenstein}}]{Grass2013}%
  \BibitemOpen
  \bibfield  {author} {\bibinfo {author} {\bibfnamefont {T.}~\bibnamefont
  {Grass}}, \bibinfo {author} {\bibfnamefont {B.}~\bibnamefont {Julia-Diaz}},
  \bibinfo {author} {\bibfnamefont {M.}~\bibnamefont {Kus}}, \ and\ \bibinfo
  {author} {\bibfnamefont {M.}~\bibnamefont {Lewenstein}},\ }\href@noop {}
  {\bibfield  {journal} {\bibinfo  {journal} {Phys. Rev. Lett.}\ }\textbf
  {\bibinfo {volume} {111}},\ \bibinfo {pages} {090404} (\bibinfo {year}
  {2013})}\BibitemShut {NoStop}%
\bibitem [{\citenamefont {Kim}\ \emph {et~al.}(2009)\citenamefont {Kim},
  \citenamefont {Chang}, \citenamefont {Islam}, \citenamefont {Korenblit},
  \citenamefont {Duan},\ and\ \citenamefont {Monroe}}]{Kim2009}%
  \BibitemOpen
  \bibfield  {author} {\bibinfo {author} {\bibfnamefont {K.}~\bibnamefont
  {Kim}}, \bibinfo {author} {\bibfnamefont {M.-S.}\ \bibnamefont {Chang}},
  \bibinfo {author} {\bibfnamefont {R.}~\bibnamefont {Islam}}, \bibinfo
  {author} {\bibfnamefont {S.}~\bibnamefont {Korenblit}}, \bibinfo {author}
  {\bibfnamefont {L.-M.}\ \bibnamefont {Duan}}, \ and\ \bibinfo {author}
  {\bibfnamefont {C.}~\bibnamefont {Monroe}},\ }\href@noop {} {\bibfield
  {journal} {\bibinfo  {journal} {Phys. Rev. Lett.}\ }\textbf {\bibinfo
  {volume} {103}},\ \bibinfo {pages} {120502} (\bibinfo {year}
  {2009})}\BibitemShut {NoStop}%
\bibitem [{\citenamefont {James}(1998)}]{James1998}%
  \BibitemOpen
  \bibfield  {author} {\bibinfo {author} {\bibfnamefont {D.~F.~V.}\
  \bibnamefont {James}},\ }\href@noop {} {\bibfield  {journal} {\bibinfo
  {journal} {Applied Physics B}\ }\textbf {\bibinfo {volume} {66}},\ \bibinfo
  {pages} {181} (\bibinfo {year} {1998})}\BibitemShut {NoStop}%
\bibitem [{\citenamefont {Porras}\ and\ \citenamefont
  {Cirac}(2004)}]{Porras2004}%
  \BibitemOpen
  \bibfield  {author} {\bibinfo {author} {\bibfnamefont {D.}~\bibnamefont
  {Porras}}\ and\ \bibinfo {author} {\bibfnamefont {J.~I.}\ \bibnamefont
  {Cirac}},\ }\href@noop {} {\bibfield  {journal} {\bibinfo  {journal} {Phys.
  Rev. Lett.}\ }\textbf {\bibinfo {volume} {92}},\ \bibinfo {pages} {207901}
  (\bibinfo {year} {2004})}\BibitemShut {NoStop}%
\bibitem [{\citenamefont {Islam}\ \emph {et~al.}(2013)\citenamefont {Islam},
  \citenamefont {Senko}, \citenamefont {Campbell}, \citenamefont {Korenblit},
  \citenamefont {Smith}, \citenamefont {Lee}, \citenamefont {Edwards},
  \citenamefont {Wang}, \citenamefont {Freericks},\ and\ \citenamefont
  {Monroe}}]{Islam2013}%
  \BibitemOpen
  \bibfield  {author} {\bibinfo {author} {\bibfnamefont {R.}~\bibnamefont
  {Islam}}, \bibinfo {author} {\bibfnamefont {C.}~\bibnamefont {Senko}},
  \bibinfo {author} {\bibfnamefont {W.~C.}\ \bibnamefont {Campbell}}, \bibinfo
  {author} {\bibfnamefont {S.}~\bibnamefont {Korenblit}}, \bibinfo {author}
  {\bibfnamefont {J.}~\bibnamefont {Smith}}, \bibinfo {author} {\bibfnamefont
  {A.}~\bibnamefont {Lee}}, \bibinfo {author} {\bibfnamefont {E.~E.}\
  \bibnamefont {Edwards}}, \bibinfo {author} {\bibfnamefont {C.-C.~J.}\
  \bibnamefont {Wang}}, \bibinfo {author} {\bibfnamefont {J.~K.}\ \bibnamefont
  {Freericks}}, \ and\ \bibinfo {author} {\bibfnamefont {C.}~\bibnamefont
  {Monroe}},\ }\href@noop {} {\bibfield  {journal} {\bibinfo  {journal}
  {Science}\ }\textbf {\bibinfo {volume} {340}},\ \bibinfo {pages} {583}
  (\bibinfo {year} {2013})}\BibitemShut {NoStop}%
\bibitem [{\citenamefont {Molmer}\ and\ \citenamefont
  {Sorensen}(1999)}]{Molmer1999}%
  \BibitemOpen
  \bibfield  {author} {\bibinfo {author} {\bibfnamefont {K.}~\bibnamefont
  {Molmer}}\ and\ \bibinfo {author} {\bibfnamefont {A.}~\bibnamefont
  {Sorensen}},\ }\href@noop {} {\bibfield  {journal} {\bibinfo  {journal}
  {Phys. Rev. Lett.}\ }\textbf {\bibinfo {volume} {82}},\ \bibinfo {pages}
  {1835} (\bibinfo {year} {1999})}\BibitemShut {NoStop}%
\bibitem [{\citenamefont {Olmschenk}\ \emph {et~al.}(2007)\citenamefont
  {Olmschenk}, \citenamefont {Younge}, \citenamefont {Moehring}, \citenamefont
  {Matsukevich}, \citenamefont {Maunz},\ and\ \citenamefont
  {Monroe}}]{Olmschenk2007}%
  \BibitemOpen
  \bibfield  {author} {\bibinfo {author} {\bibfnamefont {S.}~\bibnamefont
  {Olmschenk}}, \bibinfo {author} {\bibfnamefont {K.~C.}\ \bibnamefont
  {Younge}}, \bibinfo {author} {\bibfnamefont {D.~L.}\ \bibnamefont
  {Moehring}}, \bibinfo {author} {\bibfnamefont {D.~N.}\ \bibnamefont
  {Matsukevich}}, \bibinfo {author} {\bibfnamefont {P.}~\bibnamefont {Maunz}},
  \ and\ \bibinfo {author} {\bibfnamefont {C.}~\bibnamefont {Monroe}},\
  }\href@noop {} {\bibfield  {journal} {\bibinfo  {journal} {Phys. Rev. A}\
  }\textbf {\bibinfo {volume} {76}},\ \bibinfo {pages} {052314} (\bibinfo
  {year} {2007})}\BibitemShut {NoStop}%
\bibitem [{\citenamefont {Collins}\ \emph {et~al.}(2002)\citenamefont
  {Collins}, \citenamefont {Gisin}, \citenamefont {Linden}, \citenamefont
  {Massar},\ and\ \citenamefont {Popescu}}]{Collins2002}%
  \BibitemOpen
  \bibfield  {author} {\bibinfo {author} {\bibfnamefont {D.}~\bibnamefont
  {Collins}}, \bibinfo {author} {\bibfnamefont {N.}~\bibnamefont {Gisin}},
  \bibinfo {author} {\bibfnamefont {N.}~\bibnamefont {Linden}}, \bibinfo
  {author} {\bibfnamefont {S.}~\bibnamefont {Massar}}, \ and\ \bibinfo {author}
  {\bibfnamefont {S.}~\bibnamefont {Popescu}},\ }\href@noop {} {\bibfield
  {journal} {\bibinfo  {journal} {Phys. Rev. Lett.}\ }\textbf {\bibinfo
  {volume} {88}},\ \bibinfo {pages} {040404} (\bibinfo {year}
  {2002})}\BibitemShut {NoStop}%
\bibitem [{\citenamefont {Sackett}\ \emph {et~al.}(2000)\citenamefont
  {Sackett}, \citenamefont {Kielpinski}, \citenamefont {King}, \citenamefont
  {Langer}, \citenamefont {Meyer}, \citenamefont {Myatt}, \citenamefont {Rowe},
  \citenamefont {Turchette}, \citenamefont {Itano}, \citenamefont {Wineland},\
  and\ \citenamefont {Monroe}}]{Sackett2000}%
  \BibitemOpen
  \bibfield  {author} {\bibinfo {author} {\bibfnamefont {C.~A.}\ \bibnamefont
  {Sackett}}, \bibinfo {author} {\bibfnamefont {D.}~\bibnamefont {Kielpinski}},
  \bibinfo {author} {\bibfnamefont {B.~E.}\ \bibnamefont {King}}, \bibinfo
  {author} {\bibfnamefont {C.}~\bibnamefont {Langer}}, \bibinfo {author}
  {\bibfnamefont {V.}~\bibnamefont {Meyer}}, \bibinfo {author} {\bibfnamefont
  {C.~J.}\ \bibnamefont {Myatt}}, \bibinfo {author} {\bibfnamefont
  {M.}~\bibnamefont {Rowe}}, \bibinfo {author} {\bibfnamefont {Q.~A.}\
  \bibnamefont {Turchette}}, \bibinfo {author} {\bibfnamefont {W.~M.}\
  \bibnamefont {Itano}}, \bibinfo {author} {\bibfnamefont {D.~J.}\ \bibnamefont
  {Wineland}}, \ and\ \bibinfo {author} {\bibfnamefont {C.}~\bibnamefont
  {Monroe}},\ }\href@noop {} {\bibfield  {journal} {\bibinfo  {journal}
  {Nature}\ }\textbf {\bibinfo {volume} {404}},\ \bibinfo {pages} {256}
  (\bibinfo {year} {2000})}\BibitemShut {NoStop}%
\bibitem [{\citenamefont {Brown}\ \emph {et~al.}(2004)\citenamefont {Brown},
  \citenamefont {Harrow},\ and\ \citenamefont {Chuang}}]{Brown2004}%
  \BibitemOpen
  \bibfield  {author} {\bibinfo {author} {\bibfnamefont {K.~R.}\ \bibnamefont
  {Brown}}, \bibinfo {author} {\bibfnamefont {A.~W.}\ \bibnamefont {Harrow}}, \
  and\ \bibinfo {author} {\bibfnamefont {I.~L.}\ \bibnamefont {Chuang}},\
  }\href@noop {} {\bibfield  {journal} {\bibinfo  {journal} {Phys. Rev. A}\
  }\textbf {\bibinfo {volume} {70}},\ \bibinfo {pages} {052318} (\bibinfo
  {year} {2004})}\BibitemShut {NoStop}%
\bibitem [{\citenamefont {Albrecht}\ \emph {et~al.}(2014)\citenamefont
  {Albrecht}, \citenamefont {Koplovitz}, \citenamefont {Retzker}, \citenamefont
  {Jelezko}, \citenamefont {Yochelis}, \citenamefont {Porath}, \citenamefont
  {Nevo}, \citenamefont {Shoseyov}, \citenamefont {Paltiel},\ and\
  \citenamefont {Plenio}}]{Albrecht2014}%
  \BibitemOpen
  \bibfield  {author} {\bibinfo {author} {\bibfnamefont {A.}~\bibnamefont
  {Albrecht}}, \bibinfo {author} {\bibfnamefont {G.}~\bibnamefont {Koplovitz}},
  \bibinfo {author} {\bibfnamefont {A.}~\bibnamefont {Retzker}}, \bibinfo
  {author} {\bibfnamefont {F.}~\bibnamefont {Jelezko}}, \bibinfo {author}
  {\bibfnamefont {S.}~\bibnamefont {Yochelis}}, \bibinfo {author}
  {\bibfnamefont {D.}~\bibnamefont {Porath}}, \bibinfo {author} {\bibfnamefont
  {Y.}~\bibnamefont {Nevo}}, \bibinfo {author} {\bibfnamefont {O.}~\bibnamefont
  {Shoseyov}}, \bibinfo {author} {\bibfnamefont {Y.}~\bibnamefont {Paltiel}}, \
  and\ \bibinfo {author} {\bibfnamefont {M.~B.}\ \bibnamefont {Plenio}},\
  }\href@noop {} {\bibfield  {journal} {\bibinfo  {journal} {New J. Phys.}\
  }\textbf {\bibinfo {volume} {16}},\ \bibinfo {pages} {093002} (\bibinfo
  {year} {2014})}\BibitemShut {NoStop}%
\bibitem [{\citenamefont {Cai}\ \emph {et~al.}(2012)\citenamefont {Cai},
  \citenamefont {Naydenov}, \citenamefont {Pfeiffer}, \citenamefont
  {McGuinness}, \citenamefont {Jahnke}, \citenamefont {Jelezko}, \citenamefont
  {Plenio},\ and\ \citenamefont {Retzker}}]{Cai2012}%
  \BibitemOpen
  \bibfield  {author} {\bibinfo {author} {\bibfnamefont {J.-M.}\ \bibnamefont
  {Cai}}, \bibinfo {author} {\bibfnamefont {B.}~\bibnamefont {Naydenov}},
  \bibinfo {author} {\bibfnamefont {R.}~\bibnamefont {Pfeiffer}}, \bibinfo
  {author} {\bibfnamefont {L.~P.}\ \bibnamefont {McGuinness}}, \bibinfo
  {author} {\bibfnamefont {K.~D.}\ \bibnamefont {Jahnke}}, \bibinfo {author}
  {\bibfnamefont {F.}~\bibnamefont {Jelezko}}, \bibinfo {author} {\bibfnamefont
  {M.~B.}\ \bibnamefont {Plenio}}, \ and\ \bibinfo {author} {\bibfnamefont
  {A.}~\bibnamefont {Retzker}},\ }\href@noop {} {\bibfield  {journal} {\bibinfo
   {journal} {New J. Phys.}\ }\textbf {\bibinfo {volume} {14}},\ \bibinfo
  {pages} {113023} (\bibinfo {year} {2012})}\BibitemShut {NoStop}%
\bibitem [{\citenamefont {Biercuk}\ \emph {et~al.}(2009)\citenamefont
  {Biercuk}, \citenamefont {Uys}, \citenamefont {VanDevender}, \citenamefont
  {Shiga}, \citenamefont {Itano},\ and\ \citenamefont
  {Bollinger}}]{Biercuk2009PRA}%
  \BibitemOpen
  \bibfield  {author} {\bibinfo {author} {\bibfnamefont {M.~J.}\ \bibnamefont
  {Biercuk}}, \bibinfo {author} {\bibfnamefont {H.}~\bibnamefont {Uys}},
  \bibinfo {author} {\bibfnamefont {A.~P.}\ \bibnamefont {VanDevender}},
  \bibinfo {author} {\bibfnamefont {N.}~\bibnamefont {Shiga}}, \bibinfo
  {author} {\bibfnamefont {W.~M.}\ \bibnamefont {Itano}}, \ and\ \bibinfo
  {author} {\bibfnamefont {J.~J.}\ \bibnamefont {Bollinger}},\ }\href@noop {}
  {\bibfield  {journal} {\bibinfo  {journal} {Phys. Rev. A}\ }\textbf {\bibinfo
  {volume} {79}},\ \bibinfo {pages} {062324} (\bibinfo {year}
  {2009})}\BibitemShut {NoStop}%
\bibitem [{\citenamefont {Murashima}\ \emph {et~al.}(2005)\citenamefont
  {Murashima}, \citenamefont {Hijii}, \citenamefont {Nomura},\ and\
  \citenamefont {Tonegawa}}]{Murashima2005}%
  \BibitemOpen
  \bibfield  {author} {\bibinfo {author} {\bibfnamefont {T.}~\bibnamefont
  {Murashima}}, \bibinfo {author} {\bibfnamefont {K.}~\bibnamefont {Hijii}},
  \bibinfo {author} {\bibfnamefont {K.}~\bibnamefont {Nomura}}, \ and\ \bibinfo
  {author} {\bibfnamefont {T.}~\bibnamefont {Tonegawa}},\ }\href@noop {}
  {\bibfield  {journal} {\bibinfo  {journal} {Journal of the Physical Society
  of Japan}\ }\textbf {\bibinfo {volume} {74}},\ \bibinfo {pages} {1544}
  (\bibinfo {year} {2005})}\BibitemShut {NoStop}%
\bibitem [{\citenamefont {Affleck}\ \emph {et~al.}(1988)\citenamefont
  {Affleck}, \citenamefont {Kennedy}, \citenamefont {Lieb},\ and\ \citenamefont
  {Tasaki}}]{Affleck1988}%
  \BibitemOpen
  \bibfield  {author} {\bibinfo {author} {\bibfnamefont {I.}~\bibnamefont
  {Affleck}}, \bibinfo {author} {\bibfnamefont {T.}~\bibnamefont {Kennedy}},
  \bibinfo {author} {\bibfnamefont {E.}~\bibnamefont {Lieb}}, \ and\ \bibinfo
  {author} {\bibfnamefont {H.}~\bibnamefont {Tasaki}},\ }\href@noop {}
  {\bibfield  {journal} {\bibinfo  {journal} {Communications in Mathematical
  Physics}\ }\textbf {\bibinfo {volume} {115}},\ \bibinfo {pages} {477}
  (\bibinfo {year} {1988})}\BibitemShut {NoStop}%
\end{thebibliography}%

\end{document}